\documentclass[twocolumn,twoside,slac_two]{revtex4}
\usepackage{graphicx}
\usepackage{fancyhdr}
\usepackage{amsmath}
\usepackage{epsfig}
\begin{document}

\title{The use of quantum-correlated \boldmath{$D^0$} decays for \boldmath{$\phi_3$} measurement}

\author{A. Bondar and A. Poluektov}
\affiliation{Budker Institute of Nuclear Physics, Novosibirsk, Russia}


\newcommand{\bdk}{$B^{\pm}\to DK^{\pm}$}
\newcommand{\bdtk}{$B^{\pm}\to \tilde{D}K^{\pm}$}
\newcommand{\bdsk}{$B^{\pm}\to D^{*}K^{\pm}$}
\newcommand{\bdstk}{$B^{\pm}\to \tilde{D}^{*}K^{\pm}$}
\newcommand{\bdgk}{$B^{\pm}\to D^{*}(D\gamma)K^{\pm}$}
\newcommand{\bdgtk}{$B^{\pm}\to \tilde{D}^{*}(D\gamma)K^{\pm}$}
\newcommand{\bdks}{$B^{\pm}\to DK^{*\pm}$}
\newcommand{\bdtks}{$B^{\pm}\to \tilde{D}K^{*\pm}$}
\newcommand{\bdksnr}{$B^{\pm}\to DK^0_S\pi^{\pm}$}
\newcommand{\bddsk}{$B^{\pm}\to D^{(*)}K^{\pm}$}
\newcommand{\bddstk}{$B^{\pm}\to \tilde{D}^{(*)}K^{\pm}$}
\newcommand{\bddsks}{$B^{\pm}\to D^{(*)}K^{(*)\pm}$}
\newcommand{\bddstks}{$B^{\pm}\to \tilde{D}^{(*)}K^{(*)\pm}$}
\newcommand{\bdpi}{$B^{\pm}\to D\pi^{\pm}$}
\newcommand{\bdtpi}{$B^{\pm}\to \tilde{D}\pi^{\pm}$}
\newcommand{\bdspi}{$B^{\pm}\to D^{*}\pi^{\pm}$}
\newcommand{\bdstpi}{$B^{\pm}\to \tilde{D}^{*}\pi^{\pm}$}
\newcommand{\bdgpi}{$B^{\pm}\to D^{*}(D\gamma)\pi^{\pm}$}
\newcommand{\bdgtpi}{$B^{\pm}\to \tilde{D}^{*}(D\gamma)\pi^{\pm}$}
\newcommand{\bddspi}{$B^{\pm}\to D^{(*)}\pi^{\pm}$}
\newcommand{\bddstpi}{$B^{\pm}\to \tilde{D}^{(*)}\pi^{\pm}$}
\newcommand{\dsdpi}{$D^{*\pm}\to D\pi^{\pm}$}
\newcommand{\dsdpis}{$D^{*\pm}\to D\pi_s^{\pm}$}
\newcommand{\dkpp}{$\overline{D}{}^0\to K^0_S\pi^+\pi^-$}
\newcommand{\dtkpp}{$\tilde{D}\to K^0_S\pi^+\pi^-$}
\newcommand{\dn}{$D^0$}
\newcommand{\dnbar}{$\overline{D}{}^0$}
\renewcommand{\arraystretch}{1.4}

\begin{abstract}
  We report the results of the Monte-Carlo study of the method to 
  determine the CKM angle $\phi_3$ using Dalitz plot analysis of $D^0$ produced 
  in \bdk\ decay. Our main goal is to find the optimal strategy for a 
  model-independent $\phi_3$ extraction. We find that the analysis using 
  decays of $CP$-tagged $D$ mesons only cannot provide a 
  completely model-independent measurement in the case of a limited data sample. 
  The procedure involving binned analysis of \bdk\ and 
  $\psi(3770)\to (K^0_S\pi^+\pi^-)_D(K^0_S\pi^+\pi^-)_D$ decays is proposed
  which, in contrast, allows not only to reach the $\phi_3$ precision 
  comparable to an unbinned model-dependent fit, but also provides an unbiased 
  measurement with currently available data. 
\end{abstract}
%

\maketitle

\section{Introduction}

A measurement of the angle $\phi_3$ ($\gamma$) of the unitarity 
triangle using Dalitz plot analysis of \dkpp\ decay 
from $B^{\pm}\to DK^{\pm}$ process, introduced by Giri {\em et al.}~\cite{giri} and the Belle 
collaboration~\cite{binp_belle} and successfully implemented by 
BaBar~\cite{babar_phi3} and Belle~\cite{belle_phi3}, presently 
offers the best constraints on this quantity. However, this technique is 
sensitive to the choice of the model used to describe the three-body $D^0$ 
decay. Currently, this uncertainty is estimated to be 
$\sim 10^{\circ}$ and due to a large statistical error does not affect the 
precision of $\phi_3$ measurement. As the amount of $B$ factory data 
increases, though, this uncertainty will become a major limitation. 
Fortunately, a model-independent approach exists (see~\cite{giri}), which 
uses the data of the $\tau$-charm factory to obtain missing information 
about the $D^0$ decay amplitude. 

In our previous study of the model-independent Dalitz analysis 
technique~\cite{phi3_modind} we have implemented a procedure proposed 
by Giri {\em et al.} that uses decays of $D$ meson in $CP$
eigenstate (we denote them as $D_{CP}$) to $K^0_S\pi^+\pi^-$. Such decay can be obtained at the 
$e^+e^-$ machine operated at the $\psi(3770)$ resonance, which decays to 
a pair of $D$ mesons. The antisymmetry of the wave function 
of the $D\overline{D}$ state induces quantum correlations between the 
decay amplitudes of two $D$ mesons. 
In particular, if one $D$ meson is reconstructed in a $CP$
eigenstate (such as $\pi^+\pi^-$ or $K^0_S\pi^0$), the other 
$D$ meson is required to have the opposite $CP$ parity. 
The procedure we have studied involves the division of the \dkpp\ 
Dalitz plots from flavor $D^0$, $D_{CP}$ and \bdk\ decay into bins.
The value of $\phi_3$ is then obtained by solving the system of equations 
that includes the numbers of events in these bins. 
We have shown that this procedure allows to measure the phase $\phi_3$ with 
the statistical precision only 30--40\% worse than in the unbinned 
model-dependent case. We did not attempt to optimize the 
precision and mainly considered a high-statistics limit with an aim 
to estimate the sensitivity of the future super-B factory. 

Decays $\psi(3770)\to D^0\overline{D}{}^0$ with both neutral $D$ mesons decaying to 
$K_S^0\pi^+\pi^-$ (we will refer to these decays as 
$(K_S^0\pi^+\pi^-)^2$) have also been shown to include the information useful 
for a model-independent $\phi_3$ measurement \cite{kppkpp}. These decays, 
together with the $CP$-tagged \dkpp\ decays, are presently available 
at the CLEO-c experiment \cite{cleoc1,cleoc4}. 
The first analyses using data collected at $\psi(3770)$ resonance involve 
$\sim 400$~pb$^{-1}$ data set, while by the end of CLEO-c operation 
the integrated luminosity at the $\psi(3770)$ will reach 
750~pb$^{-1}$~\cite{david, jim_bbcb}. This corresponds to $\sim 1000$ 
$CP$-tagged \dkpp\ events and $(K_S^0\pi^+\pi^-)^2$ events. 
The actual numbers may vary by a factor of two depending on the 
details of the particular analysis. 
In this paper, we report on studies of the model-independent 
approach with a limited statistics of both $\psi(3770)$ and $B$ data, 
using the $D_{CP}\to K_S^0\pi^+\pi^-$ and 
$(K_S^0\pi^+\pi^-)^2$ final states. The technique described can be applied 
to other three-body $D^0$ final states, such as $\pi^+\pi^-\pi^0$ state 
recently used by the BaBar collaboration for a model-dependent $\phi_3$ measurement 
\cite{babar_3pi}, or $K^0_SK^+K^-$ state. 

In Section \ref{sec_modind_basics} we remind the basic idea of the 
model-independent technique of $\phi_3$ determination and introduce the 
notation. Section \ref{sec_dcp_binned} is devoted to the binned analysis 
using $D_{CP}$ data sample; we propose a way to reach the statistical 
sensitivity comparable to the model-dependent technique and discuss the 
limitations of this approach related to a limited charm data set. 
In Section \ref{sec_kspipi}, we discuss how the $(K_S^0\pi^+\pi^-)^2$
sample can be utilized in a most efficient way, and obtain 
quantitative estimate of the statistical sensitivity of this approach. 

\section{Model-independent approach}

\label{sec_modind_basics}

The density of \dkpp\ Dalitz plot is given by the absolute value of 
the amplitude $f_D$ squared: 
\begin{equation}
  p_D=p_D(m^2_+, m^2_-)=|f_D(m^2_+, m^2_-)|^2. 
  \label{p_d}
\end{equation}
The effects of charm mixing are not included in our formulas. 
For the currently expected $\phi_3$ accuracy and present limits on 
parameters of $D^0$ mixing ($x_D, y_D\sim 0.01$ \cite{mixing_hfag}), 
these effects can be safely neglected \cite{mixing_phi3}, although 
it is possible to take them into account if they appear 
to be significant for future precision measurements. 

In the case of no $CP$-violation in $D$ decay the density of the \dn\ 
decay $\bar{p}_D$ equals
\begin{equation}
  \bar{p}_D=|\bar{f}_D|^2=p_D(m^2_-, m^2_+). 
\end{equation}
Then the density of the $D$ decay Dalitz plot from \bdk\ decay is 
expressed as
\begin{equation}
\begin{split}
  p_{B^{\pm}}=&|f_D+ r_Be^{i(\delta_B\pm\phi_3)}\bar{f}_D|^2=\\
              &p_D+r_B^2\bar{p}_D+2\sqrt{p_D\bar{p}_D}(x_{\pm}c+y_{\pm}s), 
  \label{p_b}
\end{split}
\end{equation}
where $x_{\pm},y_{\pm}$ include the value of $\phi_3$ and other related 
quantities, the ratio $r_B$ of the absolute values of interfering 
$B^+\to \overline{D}{}^0K^+$ and $B^+\to D^0K^+$ amplitudes
(or their charge-conjugate partners), and 
the strong phase difference $\delta_B$ between these amplitudes:
\begin{equation}
  x_{\pm}=r_B\cos(\delta_B\pm\phi_3); \;\;\;
  y_{\pm}=r_B\sin(\delta_B\pm\phi_3). 
\end{equation}
The functions $c$ and $s$ are the cosine and sine of the strong phase 
difference $\Delta\delta_D$ between the symmetric Dalitz plot points: 
\begin{equation}
\begin{split}
  c=&\cos(\delta_D(m^2_+,m^2_-)-\delta_D(m^2_-,m^2_+))=\cos\Delta\delta_D; \\
  s=&\sin(\delta_D(m^2_+,m^2_-)-\delta_D(m^2_-,m^2_+))=\sin\Delta\delta_D. 
\end{split}
\end{equation}
The phase difference $\Delta\delta_D$ can be obtained from the sample of 
$D$ mesons in a $CP$-eigenstate, either $CP$-even or $CP$-odd, 
decaying to $K_S^0\pi^+\pi^-$. The Dalitz plot density of such decays is 
\begin{equation}
  p_{CP}=|f_D\pm \bar{f}_D|^2=p_D+\bar{p}_D\pm 2\sqrt{p_D\bar{p}_D}c
  \label{p_cp}
\end{equation}
(the normalization is arbitrary). 

Another possibility is to use a sample where both $D$ mesons (we denote them 
as $D$ and $D'$) from the $\psi(3770)$ meson decay into the 
$K_S^0\pi^+\pi^-$ state~\cite{kppkpp}. 
Since the $\psi(3770)$ is a vector, two $D$ mesons are produced in a $P$-wave, 
and the wave function of the two mesons is antisymmetric. 
Then the four-dimensional density of two correlated Dalitz plots is
\begin{equation}
\begin{split}
  p_{\rm corr}&(m_+^2,m_-^2,m'^2_+,m'^2_-)=|f_D\bar{f}_D'-f_D'\bar{f}_D|^2=\\
     &p_D\bar{p}_D'+\bar{p}_Dp_D'-2\sqrt{p_D\bar{p}_Dp_D'\bar{p}_D'}(cc'+ss'),
  \label{p_corr}
\end{split}
\end{equation}
This decay is sensitive to both $c$ and $s$ for the price of having to deal with 
the four-dimensional phase space. 

In a real experiment, one measures scattered data rather than a 
probability density. To deal with real data, the Dalitz plot can be 
divided into bins. In what follows, we show that using 
appropriate binning, it is possible to reach the statistical 
sensitivity of $\phi_3$ measurement equivalent to the model-dependent approach.

\section{Binned analysis with $D_{CP}$ data}

\label{sec_dcp_binned}

Assume that the Dalitz plot is divided into $2\mathcal{N}$ bins symmetrically
to the exchange $m^2_-\leftrightarrow m^2_+$. The bins are denoted
by the index $i$ ranging from $-\mathcal{N}$ to $\mathcal{N}$ (excluding 0); 
the exchange $m^2_+ \leftrightarrow m^2_-$ corresponds to the exchange 
$i\leftrightarrow -i$. Then the expected number of events 
in the bins of the Dalitz plot of $D$ decay from \bdk\ is 
\begin{equation}
  \langle N_i\rangle = h_B[K_i + r_B^2K_{-i} + 2\sqrt{K_iK_{-i}}(xc_i+ys_i)], 
  \label{n_b}
\end{equation}
where $K_i$ is the number of events in the bins in the Dalitz plot 
of the $D^0$ in a flavor eigenstate, $h_B$ is the normalization
constant. Coefficients $c_i$ and $s_i$, which include the information about 
the cosine and sine of the phase difference, are given by
\begin{equation}
  c_i=\frac{\int\limits_{\mathcal{D}_i}
            \sqrt{p_D\bar{p}_D}
            \cos(\Delta\delta_D(m^2_+,m^2_-))d\mathcal{D}
            }{\sqrt{
            \int\limits_{\mathcal{D}_i}p_Dd\mathcal{D}
            \int\limits_{\mathcal{D}_i}\bar{p}_Dd\mathcal{D}
            }}, 
  \label{cs}
\end{equation}
$s_i$ is defined similarly with cosine substituted by sine. 
Here $\mathcal{D}_i$ is the bin region over which the integration is 
performed. Note that $c_i=c_{-i}$, $s_i=-s_{-i}$ and $c_i^2+s_i^2\leq 1$
(the equality $c_i^2+s_i^2=1$ being satisfied if the amplitude is constant
across the bin). 

The coefficients $K_i$ are obtained precisely from a very large sample 
of $D^0$ decays in the flavor eigenstate, which is accessible at $B$-factories. 
The expected number of events in the Dalitz plot of $D_{CP}$ decay equals to
\begin{equation}
  \langle M_i\rangle = h_{CP}[K_i + K_{-i} + 2\sqrt{K_iK_{-i}}c_i], 
\end{equation}
and thus can be used to obtain the coefficient $c_i$. As soon as the $c_i$ and 
$s_i$ coefficients are known, one can obtain $x$ and $y$ values (hence, $\phi_3$
and other related quantities) by a maximum likelihood fit using equation (\ref{n_b}). 

Note that now the quantities of interest $x$ and $y$ (and consequently $\phi_3$)
have two statistical errors: one due to a finite sample of \bdk\ data, and 
the other due to $D_{CP}\to K^0_S\pi^+\pi^-$ statistics. 
We will refer to these errors as $B$-statistical and $D_{CP}$-statistical, 
respectively. 

Obtaining $s_i$ is a major problem in this analysis. If binning is fine 
enough, so that both the phase difference $\Delta\delta_D$ and the amplitude 
$|f_D|$ remain 
constant across the area of each bin, expressions (\ref{cs}) reduce to 
$c_i=\cos(\Delta\delta_D)$ and $s_i=\sin(\Delta\delta_D)$, so $s_i$ can 
be obtained as $s_i=\pm\sqrt{1-c_i^2}$. Using this equality if the amplitude
varies will lead to the bias in the $x,y$ fit result. Since $c_i$ is obtained 
directly, and $s_i$ is overestimated by the absolute value, the bias will 
mainly affect $y$ determination, resulting in lower absolute values of $y$. 

Our studies \cite{phi3_modind} show that the use of the equality $c_i^2+s_i^2=1$ 
is satisfactory for the number of bins around 200 or more, which cannot be used 
with presently available $D_{CP}$ data. It is therefore essential to find a 
relatively coarse binning (the number of bins being 10--20) which 
a) allows to extract $s_i$ from $c_i$ with low bias, and 
b) has the sensitivity to the $\phi_3$ phase 
comparable to the unbinned model-dependent case. 

Fortunately, both the a) and b) requirements appear to be equivalent. 
To determine the $B$-statistical sensitivity of a certain binning, 
let's define a quantity $Q$ --- a ratio of a statistical sensitivity 
to that in the unbinned case. Specifically, $Q$ relates the number of 
standard deviations by which the number of events in bins is changed by 
varying parameters $x$ and $y$, to the number of standard deviations 
if the Dalitz plot is divided into infinitely small regions 
(the unbinned case):
\begin{equation}
  Q^2=\frac{\sum\limits_i
    \left(\frac{1}{\sqrt{F_i}}\frac{dF_i}{dx}\right)^2+
    \left(\frac{1}{\sqrt{F_i}}\frac{dF_i}{dy}\right)^2
  }{
    \int\limits_{\mathcal{D}}\left[
      \left(\frac{1}{\sqrt{|f_B|^2}}\frac{d|f_B|^2}{dx}\right)^2+
      \left(\frac{1}{\sqrt{|f_B|^2}}\frac{d|f_B|^2}{dy}\right)^2
    \right]\,d\mathcal{D}
  }, 
\end{equation}
where $f_B=f_D+(x+iy)\bar{f}_D$, $F_i=\int_{\mathcal{D}_i}|f_B|^2d\mathcal{D}$. 

Since the precision of $x$ and $y$ weakly depends on the values
of $x$ and $y$ \cite{phi3_modind}, we can take for simplicity $x=y=0$. 
In this case one can show that
\begin{equation}
  Q^2|_{x=y=0}=\sum\limits_i(c^2_i+s^2_i) N_i\left/\sum\limits_i N_i\right. 
\end{equation}
Therefore, the binning which satisfies $c_i^2+s^2_i=1$ ({\em i.e.}
the absence of bias if $s_i$ is calculated as $\sqrt{1-c_i^2}$) also 
has the same sensitivity as the unbinned approach ($Q=1$). The factor $Q$ 
defined this way is not necessarily the best measure of the 
binning quality (the binning with higher $Q$ can be insensitive 
to either $x$ or $y$, which is impractical from the point of measuring 
$\phi_3$), but it allows an easy calculation and correctly reproduces the 
relative quality for a number of binnings we tried in our simulation. 

The optimal binning that gives the best $\phi_3$ precision is naturally 
model-dependent, but our goal is to find the analysis procedure that should 
give an unbiased result for any reasonable variation of the $D^0$ amplitude 
({\it i.e.} the fit procedure should be model-independent). 
In our studies we use the two-body amplitude obtained in the latest 
Belle $\phi_3$ Dalitz analysis \cite{belle_phi3}. 

\begin{figure}[h]
  \epsfig{file=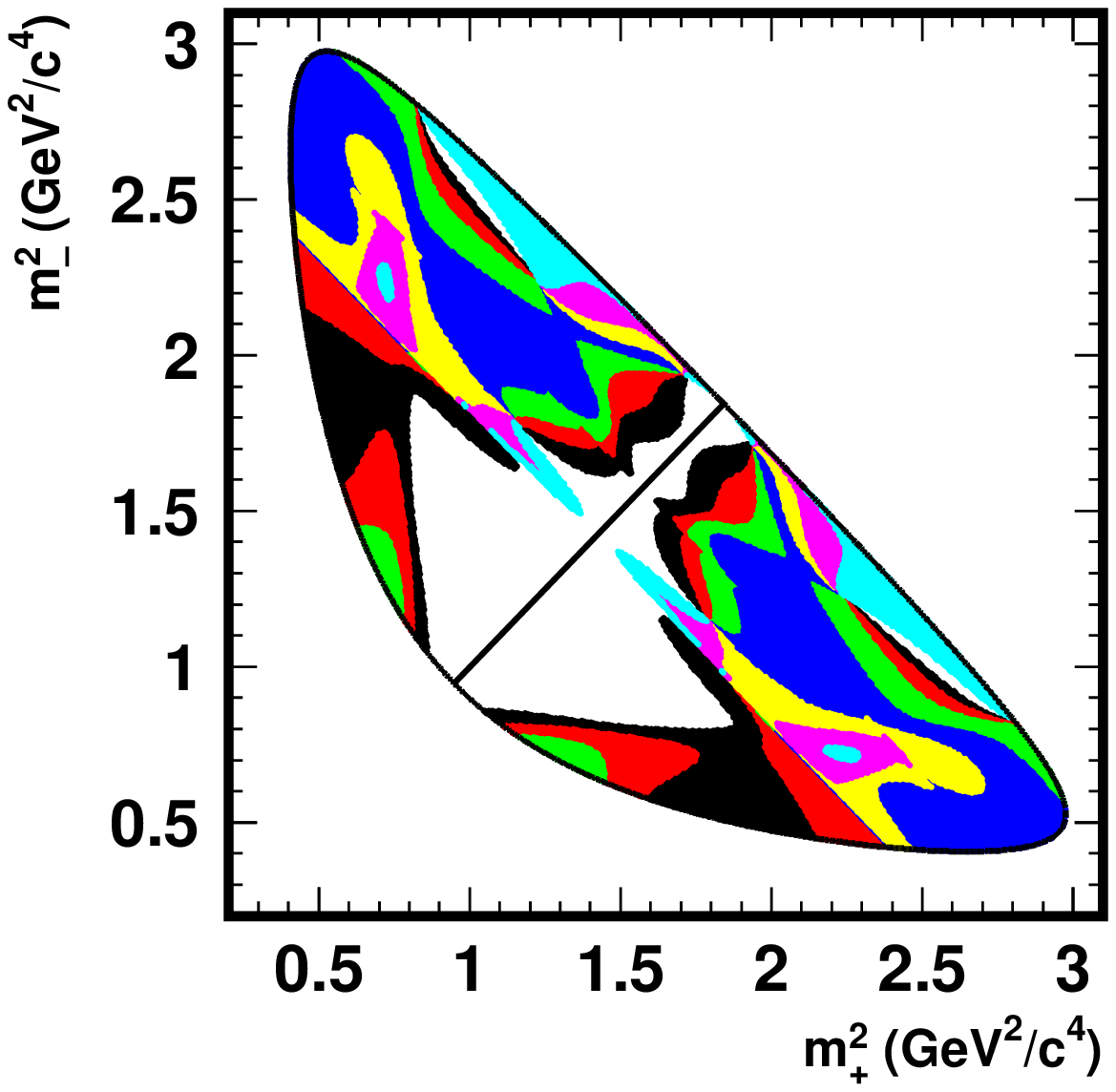, width=0.23\textwidth}
  \epsfig{file=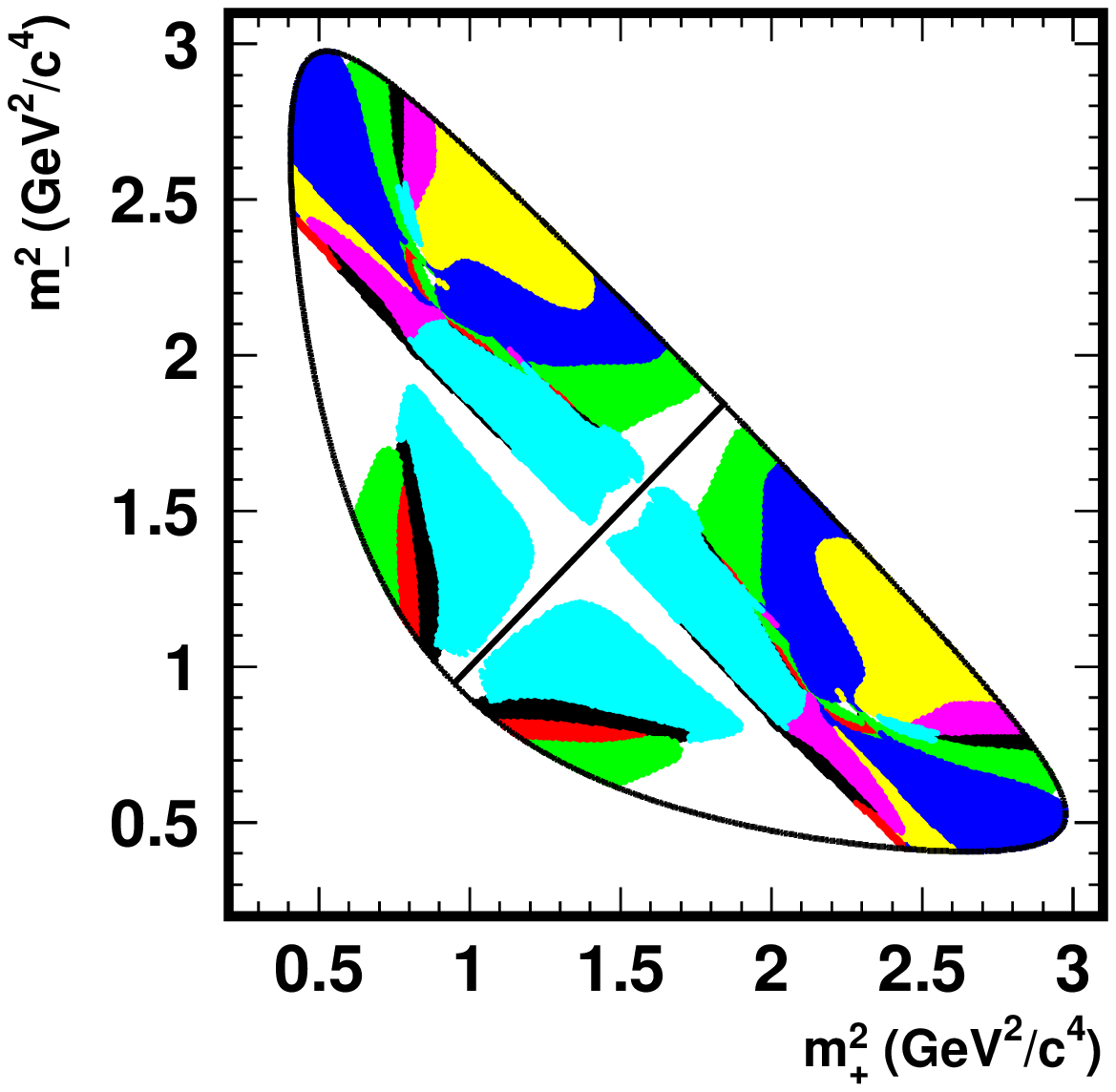, width=0.23\textwidth}
  \put(-140,90){a)}
  \put(-24,90){b)}
  \vspace{-\baselineskip}
  \caption{Divisions of the \dkpp\ Dalitz plot. Uniform binning of 
           $\Delta\delta_D$ strong phase difference with $\mathcal{N}=8$ (a), 
           and the binning obtained by variation of the latter to maximize 
           the sensitivity factor $Q$ (b). 
           }
  \label{binning}
\end{figure}

From the consideration above it is clear that a good approximation 
to the optimal binning is the one obtained from 
the uniform division of the strong phase difference $\Delta\delta_D$. 
In the half of the Dalitz plot $m^2_+<m^2_-$ ({\it i.e.} the bin index $i>0$) 
the bin $\mathcal{D}_i$ is defined by the condition
\begin{equation}
   2\pi(i-1/2)/\mathcal{N}<\Delta\delta_D(m^2_+, m^2_-)<
   2\pi(i+1/2)/\mathcal{N}, 
\end{equation}
and in the remaining part ($i<0$) the bins are defined symmetrically. 
We will refer to this binning as $\Delta\delta_D$-binning. 
As an example, such a binning with $\mathcal{N}=8$ is shown in 
Fig.~\ref{binning}~(a). Although the phase difference variation across the 
bin is small by definition, the absolute value of the amplitude can 
vary significantly, so the condition $c^2_i+s^2_i=1$ is not satisfied 
exactly. The values of $c_i$ and $s_i$ in this binning are shown in 
Fig.~\ref{cspic}c) with crosses. 

Figure~\ref{binning}~(b) shows the division with $\mathcal{N}=8$ obtained by 
continuous variation of the $\Delta\delta_D$-binning to maximize the factor $Q$. 
The sensitivity factor $Q$ increases to 0.89 compared to 0.79 for 
$\Delta\delta_D$-binning. 

\begin{table*}
  \caption{Statistical precision of $(x,y)$ determination using 
           different binnings and with an unbinned approach. The errors
           correspond to 1000 events in both the $B$ and 
           $D_{CP}$ ($(K_S^0\pi^+\pi^-)^2$) samples. 
           The $D^0$ amplitude used is the result of the Belle 
           analysis~\cite{belle_phi3}. }
  \label{stat}
  \centering
  \begin{tabular}{|l||c||c|c||c|c||cc|} 
    \hline
                                &   & \multicolumn{2}{|c||}{$B$-stat. err.} & 
                                      \multicolumn{2}{|c||}{$D_{CP}$-stat. err.} & 
                                      \multicolumn{2}{|c|}{$(K_S^0\pi^+\pi^-)^2$-stat.
                                      err.} \\ 
                                      \cline{3-8}
    Binning                     & Q & $\sigma_{x}$ & $\sigma_{y}$ & $\sigma_{x}$ & $\sigma_{y}$ & $\sigma_{x}$ & $\sigma_{y}$ \\ 
    \hline
    $\mathcal{N}=8$ (uniform)            & 0.57 & 0.033 & 0.060 & 0.005 & 0.010 & 0.015 & 0.032 \\
    $\mathcal{N}=8$ ($\Delta\delta_D$)   & 0.79 & 0.027 & 0.037 & 0.004 & 0.007 & 0.005 & 0.010 \\
    $\mathcal{N}=8$ (optimal)            & 0.89 & 0.023 & 0.032 & 0.006 & 0.011 & 0.008 & 0.011 \\
    $\mathcal{N}=19$ (uniform)           & 0.69 & 0.027 & 0.055 & 0.004 & 0.011 & 0.013 & 0.019 \\
    $\mathcal{N}=20$ ($\Delta\delta_D$)  & 0.82 & 0.027 & 0.035 & 0.005 & 0.007 & 0.004 & 0.008 \\
    $\mathcal{N}=20$ (optimal)           & 0.96 & 0.022 & 0.029 & 0.008 & 0.011 & 0.004 & 0.010 \\
    \hline
    Unbinned                    & -    & 0.021 & 0.028 & -      & -      &      - &      - \\ 
    \hline
  \end{tabular}
\end{table*}

We perform a toy MC simulation to study the statistical sensitivity of the 
different binning options. We use the amplitude from the Belle 
analysis~\cite{belle_phi3}
to generate decays of flavor $D^0$, $D_{CP}$, and $D$ from \bdk\ decay
to the $K^0_S\pi^+\pi^-$ final state according to the probability density given by 
(\ref{p_d}), (\ref{p_cp}) and (\ref{p_b}), respectively. 
In the present study we use the errors of parameters $x$ and $y$ 
rather than $\phi_3$ as a measure of the statistical power since they are 
nearly independent of the actual values of $\phi_3$, strong phase $\delta$ and 
amplitude ratio $r_B$. The error of $\phi_3$ can be obtained from these
numbers given the value of $r_B$. 
To obtain the $B$-statistical error we use a large 
number of $D^0$ and $D_{CP}$ decays, while the generated number of $D$ decays 
from the \bdk\ process ranges from $10^2$ to $10^5$. For each number of $B$ decay 
events, 100 samples are generated, and the statistical errors of 
$x$ and $y$ are obtained from the spread of the fitted values. 
A study of the error due to $D_{CP}$ statistics is performed similarly, 
with a large number of $B$ decays, and the statistics of $D_{CP}$ decays
varied. Both errors are checked to satisfy the square root scaling. 

The binning options used are $\Delta\delta_D$-binning with $\mathcal{N}=8$ 
and $\mathcal{N}=20$, 
as well as ``optimal" binnings with maximized $Q$ obtained from these two
with a smooth variation of the bin shape. 
For comparison, we use the binnings with the uniform division into rectangular
bins (with $\mathcal{N}=8$ and $\mathcal{N}=19$ in the allowed phase space, 
the ones which are denoted as 3x3 and 5x5 in~\cite{phi3_modind}). 

The $B$- and $D_{CP}$-statistical precision of different binning options, 
recalculated to 1000 events of both $B$ and $D_{CP}$ samples, 
as well as their calculated values of the factor $Q$, 
are shown in Table~\ref{stat}. 
The factor $Q$ reproduces the 
ratio of the values $\sqrt{1/\sigma_x^2+1/\sigma_y^2}$ for the binned 
and unbinned approaches with the precision of 1--2\%. 
Note that the ``optimal" binning with $\mathcal{N}=20$ offers the 
$B$-statistical sensitivity only 4\% worse than an unbinned technique. 
While the binning with 
maximized $Q$ offers better $B$-statistical 
sensitivity, the best $D_{CP}$-statistical precision 
of the options we have studied is reached for the 
$\Delta\delta_D$-binning. However, for the expected amount of experimental 
data of $B$ and $D_{CP}$ decays the $B$-statistical error dominates, 
therefore, slightly worse precision due to $D_{CP}$ statistics does 
not affect significantly the total precision. 

\begin{figure}
  \epsfig{file=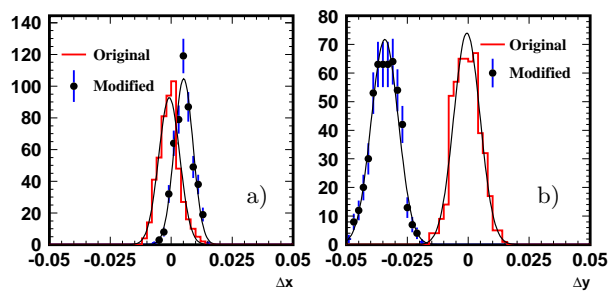, width=0.5\textwidth}
  \put(-140,40){a)}
  \put(-30,40){b)}
  \vspace{-\baselineskip}
  \caption{
            Toy MC study of the analysis using $D_{CP}$ data. 
            Difference between the fitted and generated 
            (a) $x$ and (b) $y$ values. Result of the 
            toy MC study with $\Delta\delta_D$ binning, 
            $10^5$ $B$ decays and $10^4$ $D_{CP}$ decays. 
            Histogram shows the fit result with the same $D^0$ decay 
            amplitude used for event generation and binning, 
            the points with the error bars show the case with 
            different amplitudes. 
           }
  \label{dcppic}
\end{figure}

We have considered the choice of the optimal binning only from the 
point of statistical power. However, the conditions to satisfy low model 
dependence are quite different. Since the bins in the binning options 
we have considered are sufficiently large, the requirement that the phase 
does not change over the bin area is a strong model assumption. 
We have performed toy MC simulation to study the model dependence. 
While the binning was kept the same as in the statistical power study
(based on the phase difference from the default $D^0$ amplitude), 
the amplitude used to generate $D^0$, $D_{CP}$ and \bdk\ decays
was altered in the same way as in the Belle study of the model dependence 
in the unbinned analysis~\cite{belle_phi3}. 
As a result, the same bias of $\Delta\phi_3\sim 10^{\circ}$ is observed
as in unbinned analysis. The magnitude of the bias in $x$ and $y$ 
(for initial $x=0$, $y=0.1$) is demonstrated in Fig.~\ref{dcppic}. 
This bias is apparently caused by a fixed 
relation between the $c_i$ and $s_i$, and it affects mainly the 
$y$ variable. 

In a real analysis, one can control the model error by testing 
if the amplitude used to define binning is compatible with the observed 
$D_{CP}$ data. This can be done, {\em e.g.,} by dividing each bin and 
comparing calculated values of $c_i$ in its parts, or by comparing the 
expected and observed numbers of events in each bin. The first results 
by the CLEO-c collaboration are available \cite{cleo_charm07} that 
show good agreement of experimental data with $c_i$ calculated from 
two-body amplitude for $\Delta\delta_D$-binning. 

We conclude that the method of $\phi_3$ determination
using only $D_{CP}$ data is only asymptotically 
model-in\-de\-pen\-dent, since for any finite bin size the calculation 
of $s_i$ is done using model assumptions of the $\Delta\delta_D$
variations across the bin. Increasing the $D_{CP}$
data set, however, allows to apply a finer binning and therefore 
reduce the model error due to the variation of the phase difference. 

\section{Binned analysis with correlated $D^0\to K_S^0\pi^+\pi^-$ data}

\label{sec_kspipi}

The use of $\psi(3770)$ decays where both neutral $D$ mesons decay to the
$K^0_S\pi^+\pi^-$ state allows to significantly increase the amount of data
useful to extract phase information in $D^0$ decay. It is also possible 
to detect events of $\psi(3770)\to (K^0_S\pi^+\pi^-)_D (K^0_L\pi^+\pi^-)_D$, 
where $K^0_L$ is not reconstructed, and its momentum is obtained from kinematic 
constraints. The number of these events is approximately twice that of 
$(K^0_S\pi^+\pi^-)^2$. However, it is impossible to 
simply combine these samples since the phases of the doubly Cabibbo-suppressed 
components in $\overline{D}{}^0\to K_S^0\pi^+\pi^-$ and 
$\overline{D}{}^0\to K_L^0\pi^+\pi^-$
amplitudes are opposite. In the analysis of $B$ data only $K_S^0\pi^+\pi^-$ 
state can be used, but it is possible to utilize $K^0_L\pi^+\pi^-$ 
data to better constrain the \dkpp\ amplitude using model assumptions based on
$SU(3)$ symmetry~\cite{cleo_charm07}. In what follows, we will consider the 
use of $K^0_S\pi^+\pi^-$ data only. 

In the case of a binned analysis, the number of events in the region of the
$(K^0_S\pi^+\pi^-)^2$ phase space is 
\begin{equation}
\begin{split}
  \langle M\rangle_{ij} = h_{\rm corr}[&K_i K_{-j} + K_{-i} K_j - \\
    &2\sqrt{K_iK_{-i}K_jK_{-j}}(c_i c_j + s_i s_j)]. 
\end{split}
\end{equation}
Here two indices correspond to two $D$ mesons from $\psi(3770)$ decay. 
It is logical to use the same binning as in the case of $D_{CP}$
statistics to improve the precision of the determination of 
$c_i$ coefficients, and to obtain $s_i$ from data without model 
assumptions, contrary to $D_{CP}$ case. Note that in the case of 
using $(K^0_S\pi^+\pi^-)^2$ decays, the parameters $c_i$ and $s_i$
are treated as independent variables. The obvious advantage 
of this approach is its being unbiased for any finite 
$(K_S^0\pi^+\pi^-)^2$ statistics (not only asymptotically as in the case of 
$D_{CP}$ data).

Note that in contrast to $D_{CP}$ analysis, where the sign of $s_i$
in each bin is undetermined and has to be fixed using model assumptions, 
$(K^0_S\pi^+\pi^-)^2$ analysis has only a four-fold ambiguity: change of 
the sign of all $c_i$ or all $s_i$. In combination with $D_{CP}$ analysis, 
where the sign of $c_i$ is fixed, this ambiguity reduces to only two-fold.  
One of the two solutions can be chosen based on a weak model assumption
(incorrect $s_i$ sign corresponds to complex-conjugate $D$ decay
amplitude, which violates a causality requirement when parameterized with 
the Breit-Wigner amplitudes). 

The coefficients $c_i$, $s_i$ can be obtained by minimizing the negative 
logarithmic likelihood function
\begin{equation}
  -2\log\mathcal{L}=-2\sum\limits_{i,j}\log P(M_{ij}, \langle M\rangle_{ij}), 
  \label{corr_lh}
\end{equation}
where $P(M,\langle M\rangle)$ is the Poisson probability to get $M$ events with the 
expected number of $\langle M\rangle$ events. 

The number of bins in the 4-dimensional phase space is $4\mathcal{N}^2$
rather than $2\mathcal{N}$ in the $D_{CP}$ case. Since the expected number of 
events in correlated $K^0_S\pi^+\pi^-$ data is of the same order as for $D_{CP}$, 
the bins will be much less populated. This, however, does not affect the 
precision of the $c_i$, $s_i$ determination since 
the number of free parameters is the same and each of the parameters 
is constrained by many bins. 

The coefficients $c_i$, $s_i$ obtained this way can then be used to 
constrain $x$, $y$ with the maximum likelihood fit of the $B$ decay data 
using Eq.~\ref{n_b}. To correctly account for the errors of the $c_i$, $s_i$
determination, this likelihood should include distributions of these 
quantities, in addition to Poisson fluctuations in the $B$ data bins. A more 
convenient way is to use the common likelihood function, 
covering both $B$ and $K_S^0\pi^+\pi^-$ data: 
\begin{equation}
\begin{split}
  -2\log\mathcal{L}=&-2\sum\limits_{i,j}\log P(M_{ij}, \langle M\rangle_{ij})\\
                    &-2\sum\limits_{i}\log P(N_i, \langle N\rangle_{i}), 
  \label{comb_lh}
\end{split}
\end{equation}
with $x$, $y$, $h_B$, $h_{corr}$, $c_i$ and $s_i$ as free parameters. 
This approach is also more optimal in the case of large $B$ data sample, 
since it imposes additional constraints on $c_i$, $s_i$ values. 

The toy MC simulation was performed to study the procedure described above. 
Using the amplitude from the Belle analysis~\cite{belle_phi3}, we generate a large number 
of $D^0\to K^0_S\pi^+\pi^-$ decays and several sets of $(K^0_S\pi^+\pi^-)^2$ decays 
(according to the probability density given by (\ref{p_corr}))
and $B$ decays (\ref{p_b}). We use the same binning options as in 
the $D_{CP}$ study. The combined negative likelihood 
(\ref{comb_lh}) is minimized in the fit to each toy MC sample. 
We constrain $c^2_i+s^2_i<1$ in the fit to avoid entering an unphysical 
region with a negative number of events in the bin. For low number of 
$(K^0_S\pi^+\pi^-)^2$ decays this constraint introduces asymmetric
tails in the $x,y$ distributions. For $10^3$ events and more this 
asymmetry becomes negligible. Since the number 
of $(K^0_S\pi^+\pi^-)^2$ decays we expect in the experiment is of the order 
of $10^3$, we do not expect this effect to cause a significant problem. 

The number of $(K^0_S\pi^+\pi^-)^2$ and $B$ decays in our study of 
statistical sensitivity ranges from $10^3$ to 
$10^5$. The errors of $x$ and $y$ parameters are calculated from the spread 
of the fitted values. 
If the number of $(K^0_S\pi^+\pi^-)^2$ decays is comparable or larger than 
the number of $B$ decays, the $x$ and $y$ errors can be represented as 
quadratic sums of two errors, each scaled as a square root of 
$(K^0_S\pi^+\pi^-)^2$ and $B$ statistics, respectively. However if the number of 
$B$ decays is large, the errors of $c_i$ and $s_i$ depend also on 
$B$ decay statistics, so separating the total error into $B$- and 
$(K^0_S\pi^+\pi^-)^2$-statistical errors becomes impossible. 

\begin{figure}
  \epsfig{file=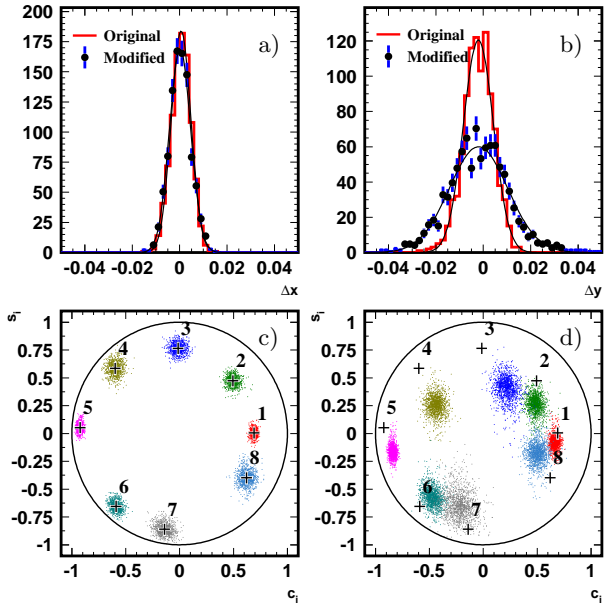, width=0.5\textwidth}
  \put(-140,215){a)}
  \put(-26,215){b)}
  \put(-140,104){c)}
  \put(-26,104){d)}
  \vspace{-\baselineskip}
  \caption{Toy MC study of the analysis using $(K_S^0\pi^+\pi^-)^2$ data. 
            Top line: difference between the fitted and generated 
            (a) $x$ and (b) $y$ values. Result of the 
            toy MC study with $\Delta\delta_D$ binning, 
            $10^5$ $B$ decays and $10^4$ $(K_S^0\pi^+\pi^-)^2$ decays. 
            The histogram shows the fit result with the same $D^0$ decay 
            amplitude used for event generation and binning, 
            the points with the error bars show the case with 
            different amplitudes. 
            Bottom line: coefficients $c_i$, $s_i$ obtained in the 
            fit to toy MC sample. Different colors correspond to different bins. 
            Cases with the same amplitude (c) and different amplitudes
            (d) used for event generation and binning. 
           }
  \label{cspic}
\end{figure}

The best $(K_S^0\pi^+\pi^-)^2$-statistical error is obtained for $\Delta\delta_D$-binning
and recalculated to 1000 events yields $\sigma_x=0.005$, $\sigma_y=0.010$, which is 
only slightly worse than the error obtained with the same amount of $D_{CP}$ data
(see Table~\ref{stat} for comparison). 
We also check that significant change of the model used to define the 
binning does not lead to the systematic bias (although it does decrease 
the statistical precision). Figure~\ref{cspic} demonstrates the precision 
of the determination of $c_i$, $s_i$ coefficients in our toy MC study and the 
absence of the systematic bias for both $x$ and $y$ when the model is varied. 

The numbers of $(K_S^0\pi^+\pi^-)^2$ and $D_{CP}$ decays in the 
$\psi(3770)$ data 
are comparable, and so are the statistical errors due to the $\psi(3770)$ data sample 
for the two approaches. However, the approach based on $(K_S^0\pi^+\pi^-)^2$ 
data allows to extract both $c_i$ and $s_i$ without additional model 
uncertainties, so it can be 
used to check the validity of the constraint $c^2_i+s^2_i=1$ and therefore 
to test the sensitivity of the particular binning. 
The same binning can be used in both $(K_S^0\pi^+\pi^-)^2$ and $D_{CP}$ 
approaches, therefore improving the accuracy of the $c_i$ determination. 
Technically it can be done in a straightforward way by adding the third
term related to the number of $D_{CP}$ decays into the likelihood (\ref{comb_lh}). 

\section{Conclusion}

We have studied the model-independent approach to $\phi_3$ measurement 
using $B^{\pm}\to DK^{\pm}$ decays 
with the neutral $D$ decaying to $K^0_S\pi^+\pi^-$. 
The analysis of $\psi(3770)\to D\bar{D}$ data allows to extract
the information about the strong phase in the \dkpp\ decay, 
whereas this phase is fixed by model assumptions in a model-dependent technique.
We consider the case with a limited $\psi(3770)\to D\bar{D}$ data 
sample which will be available from CLEO-c in the near future. 

In the binned analysis, we propose a way to obtain the binning that 
offers an optimal statistical precision (close to the precision of 
an unbinned approach). 
Two different strategies of the binned analysis are considered: 
using the $D_{CP}\to K^0_S\pi^+\pi^-$ data sample, and using decays of 
$\psi(3770)$ to $(K_S^0\pi^+\pi^-)_D (K_S^0\pi^+\pi^-)_D$. 
The strategy using $D_{CP}$ decays alone cannot offer a completely 
model-independent measurement: it provides only the information 
about $c_i$ coefficients, while $s_i$ for low $D_{CP}$ statistics 
has to be fixed using model assumptions. However, as the $D_{CP}$
data sample increases, model-independence can be reached by reducing 
the bin size. The strategy using the 
$\psi(3770)\to(K_S^0\pi^+\pi^-)_D (K_S^0\pi^+\pi^-)_D$ sample, in contrast, 
allows to obtain not only $c_i$ coefficients 
with an accuracy comparable to $D_{CP}$ approach, but also $s_i$ in 
a model-independent way. Both strategies can use the same binning of the 
\dkpp\ Dalitz plot and therefore can be used in combination to 
improve the accuracy due to $\psi(3770)$ statistics. 

The expected sensitivity to $\phi_3$ is obtained based on the 
two-body \dkpp\ decay amplitude measured by Belle~\cite{belle_phi3}. 
For the \mbox{CLEO-c} statistics of 750 pb$^{-1}$ 
($\sim 1000$ $D_{CP}$ and $(K_S^0\pi^+\pi^-)^2$ events) the expected errors 
of the parameters $x$ and $y$ due to $\psi(3770)$ statistics are 
of the order of 0.01. For $r_B=0.1$ it gives the $\phi_3$ precision 
$\sigma_{\phi_3}=\sigma_{x,y}/(\sqrt{2}r_B)\simeq 5^{\circ}$, 
which is well below the expected error due to current $B$ data sample
(the total integrated luminosity of the two $B$-factories, BaBar and Belle, 
slightly exceeds 1~ab$^{-1}$, which corresponds to $\sim 1000$ \bdk\ 
decays and $\phi_3$ precision of about $20^{\circ}$ for $r_B=0.1$). 
Further improvement of $\phi_3$ precision at the Super B factory \cite{superb}
and LHCb \cite{lhcb} will require a larger charm 
dataset, which can be provided by the BES-III experiment~\cite{bes,bes2}. 

In our study, we did not consider the experimental systematic uncertainties, 
{\em e.g.} due to imperfect knowledge of the detection efficiency or 
background composition. We believe these issues can be addressed in a 
similar manner as in already completed model-dependent analyses. 

\section{Acknowledgments}

We would like to thank David Asner, Tim Gershon, Jure Zupan and 
Simon Eidelman for fruitful discussions and suggestions on improving 
the paper.

\end{document}